\documentclass[suppldata]{interact}

\usepackage{epstopdf}
\usepackage{subfigure}

\usepackage{natbib}

\usepackage{tikz}
\newcommand{\sqdiamond}[1][fill=black]{\tikz [x=1.2ex,y=1.2ex,line width=.1ex,line join=round, yshift=-0.285ex] \draw  [#1]  (0,.5) -- (.5,1) -- (1,.5) -- (.5,0) -- (0,.5) -- cycle;}%

\usepackage{amsmath}
\usepackage{amssymb}
\usepackage{wasysym}
\usepackage{color}
\definecolor{myred}{rgb}{1,0,0}
\newcommand{\rev}[1]{#1}
\newcommand{\revII}[1]{#1}
\newcommand{\Diff}[2]{\frac{\mathrm{D} #1}{\mathrm{D} #2}}

\newcommand{\vect}[1]{\boldsymbol{#1}}
\newcommand\Tstrut{\rule{0pt}{2.6ex}}         
\newcommand\Bstrut{\rule[-0.9ex]{0pt}{0pt}}   
\begin{document}
	
\title{Treatment of solid objects in the Pencil Code using an immersed boundary method and overset grids}

\author{
	\name{J\o rgen R. Aarnes\textsuperscript{a}\thanks{CONTACT J\o rgen.~R. Aarnes. Email: jorgen.r.aarnes@ntnu.no; Tai Jin. Email: tai.jin@ucl.ac.uk}, Tai Jin\textsuperscript{b,c}, Chaoli Mao\textsuperscript{c}, Nils E.~L. Haugen\textsuperscript{a,d}, Kun Luo\textsuperscript{c},  Helge I. Andersson\textsuperscript{a}}
	\affil{\textsuperscript{a}Department of Energy and Process Engineering, Norwegian University of Science and Technology, N-7491 Trondheim, Norway; 
	\textsuperscript{b}Department of Mechanical Engineering, University College London, London WC1E 7JE, United Kingdom; \textsuperscript{c}State Key Laboratory of Clean Energy Utilization, Zhejiang University, Hangzhou, 310027, P.~R.~China;
	\textsuperscript{d}SINTEF Energy Research, N-7465 Trondheim, Norway}
}

\maketitle

	\begin{abstract}
	Two methods for solid body representation in flow simulations available in the Pencil Code \rev{are} the immersed boundary method and overset grids. These methods are quite different in terms of computational cost, flexibility and numerical accuracy. We present here an investigation of the use of the different methods with the purpose of assessing their strengths and weaknesses. At present, the overset grid method in the Pencil Code can only be used for representing cylinders in the flow. For this task it surpasses the immersed boundary method in yielding highly accurate solutions at moderate computational costs. This is partly due to local grid stretching and a body-conformal grid, and partly due to the possibility of working with local time step restrictions on different grids. The immersed boundary method makes up the lack of computational efficiency with flexibility in regards to application to complex geometries, due to a recent extension of the method that allows our implementation of it to represent arbitrarily shaped objects in the flow.
	\end{abstract}
		
	
	\section{Introduction}
	Fluid flow in a domain that contains an immersed solid object is a common case in computational fluid dynamics. Obstructions in the flow include (but are not limited to) cylinders, spheres, flat plates, rectangular or elliptical cylinders and spheroids, triangles, and complex geometries made out of a combination of these. Finding a method to represent such objects in the best possible way in simulations is not a trivial task, and the method used is often chosen specifically to the problem at hand. 
	
	For many generic shapes, such as cylinders, spheres, plates, etc., body-fitted structured meshes are commonly used to represent the object(s) in the flow. Body-fitted structured meshes conform to the object(s) in the flow domain and to the domain's other physical boundaries (inlet, outlet, walls, etc.). Depending on the flow domain and object in the flow, this may require a deformation of the grid to conform to domain boundaries, in addition to the mapping procedures to map the grid in the flow domain to a simple computational domain. This may result in a grid with unnecessary local variations of the grid (e.g., a grid that is denser than necessary in certain areas of the domain) and time consuming grid generation \citep{versteeg2007introduction}. A popular alternative to such meshes, particularly when the shape of the flow domain or objects in the flow domain is more complex, is unstructured meshes. Unstructured meshes provide the highest flexibility in grid adaptation to a particular flow geometry, and is a good alternative for complex geometries when finite-volume or finite-element formulations of the governing equations are used \citep{mavriplis1997unstructured}. Disadvantages of such grids are much larger storage requirements than for structured grids \citep{pletcher2012computational}, the need for intricate mesh generation techniques \citep{owen1998survey}, and the difficulty in achieving high-order of accuracy.

	By other choices of grid methods, the object(s) in the flow and the flow domain can be represented without the grid conforming to the object(s). Typically this is done by using a Cartesian grid, with a modification in either the flow equations or the grid cells in the immediate vicinity of the solid object(s). Popular methods of this type include immersed boundary methods (IBMs) \citep{Peskin1972,peskin2002,Mittal2005} and cut-cell methods \citep{quirk1994alternative,causon2000calculation,gunther2011cartesian}. These methods differ in that the IBM uses a Cartesian grid in the entire flow domain, while in cut-cell methods grid cells are `cut' near the objects or domain boundaries that do not conform to the grids, and the flow equations are solved on the new, modified cells \citep{ingram2003developments}. Due to this cell cutting, care must be taken such that the cut cells do not become too small, since this may be a potential source of numerical instabilities. 
	
	For the IBM, rather than modifying the grid cells near the solid object, the boundary conditions on the solid are imposed directly in the flow equations. This is done either by a continuous or a discrete forcing technique. In both cases a body-force, present due to non-conforming boundaries in the flow, is introduced in the Navier-Stokes equations. This is done either before discretization (continuous forcing) or after (direct forcing) \citep{Mittal2005}. The latter of these is the preferred method for IBM used to represent rigid boundaries. One development of the discrete forcing method is to treat the immersed boundary as a sharp interface, and to impose the boundary conditions directly. This is done by using a combination of ghost-points inside the solid and mirror/image-points in the flow domain (set by interpolation) to reconstruct the solid \citep{tseng2003ghost,berthelsen2008local}. An advantage with this approach is that the boundary conditions are handled without any added force in the flow equations, hence, the method can easily be implemented in an existing flow solver. The disadvantage is the accuracy reduction in the vicinity of the surface, although recent developments show that some of the challenges related to high-order accurate reconstructions of velocities near the surface can be overcome \citep{seo2011high,xia2014ghost}. Further, \rev{finite-difference IBMs are, in general, not mass conserving. Finite-volume approaches with cut-cell methodology is appropriate if mass and momentum conservation needs to be guaranteed~\citep{Mittal2005}.}
	
	About ten years after the emergence of the first IBM, a new method was proposed to represent solids in the flow by using several grids overset one another \citep{Steger1983, Benek1985}. Such overset grid methods (often called Chimera methods) employ body-conformal grids at the boundaries of objects in the flow, but the grids do not extend to the physical boundaries of the domain. Rather, a background grid (typically uniform Cartesian) is used, and updated flow information of overlapping grid regions is communicated between grids at every time step. 
	\rev{Note that special overset grid methods without background grids exist, like yin-yang grids where  two identical component grids are used to cover a spherical surface, thus avoiding very small grid cells close to poles of the spherical geometry \citep{kageyama2004yin}.}
	
	The flow domain \rev{resolved with overset grids} may contain a single grid overlapping another, or several grids overlapping necessitating a priority of communication and computation of solutions of the different grids \citep{Steger1987,Chesshire1990}. For complex configurations, this may require extensive preprocessing for fixed objects \citep{suhs2002pegasus} or intricate grid handling at run time for moving bodies \citep{noack2005suggar}. 
	Overset grid methods are, in general, not mass conserving, since interpolation is necessary between grids overset one another (although exceptions do exist, for finite-volume implementations of overset grids, see \cite{Part-Enander1994} and \cite{Zang1995}). The interpolation is done from donor-points on one grid to fringe-points on another. Many different interpolation procedures have been explored for this purpose, and several studies have found that using high-order interpolation between grids is beneficial in regards to the overall accuracy and stability of flow computations \citep{Sherer2005,Chicheportiche2012,Volkner2017}. 
	
	
	In the high-order compressible flow solver known as the Pencil Code \citep{pencilcode}, solid objects in the flow can be represented by different schemes.
	This makes it possible to compare \rev{different} surface representations not only for the same flow problems, but for simulations where the same finite-difference discretization, time integration, communication procedures, etc, are used. The purpose of this paper is to perform such a comparison 
	\revII{for solids represented by a ghost-points IBM and overset grids.} 
	\revII{The performance of these surface representations is assessed} 
	in terms of computational cost and accuracy for a common benchmarking case. The flow case used is the frequently appearing fluid mechanics problem of flow past a circular cylinder. Further, we wish to shed light on an advantage of the IBM implementation in the Pencil Code by simulating flow past a complex geometry. The complex geometry used as an example case is a combination of a semi-circular and a semi-elliptical cylinder.
	
	The structure of the paper is the following: In Section \ref{sec:method} the governing flow equations and the two methods for solid object representation are described, with details on their implementation in the Pencil Code included. Performance of the different methods for the flow past a cylinder in both the steady regime and the unsteady vortex shedding regime is compared in Section \ref{sec:benchmarking}. Following this, we elaborate on the extension of the IBM to complex geometries in Section \ref{sec:complex}, before concluding remarks are made in Section \ref{sec:conclusion}.
	
	\section{Methodology {\label{sec:method}}}
		\subsection{Governing equations {\label{subsec:flowEq}}}
	The governing equations of the flow are the continuity equation:
	
	\begin{equation}
		\label{eq:continuity}
		\Diff{\rho}{t} + \rho \vect{\nabla}\cdot\vect{u} = 0 \, ,
	\end{equation}
	and the momentum equation:
	\begin{equation}
		\label{eq:mom}
		\rho \Diff{\vect{u}}{t} = -\vect{\nabla} p + \vect{\nabla} \cdot \left( 2 \mu \vect{S} \right) \, ,
	\end{equation}
	where $ \rho $, $ t $, $ \vect{u} $, $ p $ and $\mu$ are the density, time, velocity vector, pressure, and dynamic viscosity ($ \mu=\rho\nu $, with kinematic viscosity $ \nu $), respectively
	\rev{, and
	\begin{equation}
	\Diff{}{t} = \frac{\partial}{\partial t}+\vect{u}\cdot \vect{\nabla} \, ,
	\end{equation}
	is the material derivative operator.} 
	The compressible rate of strain tensor $ \vect{S} $ is given by:
	\begin{equation}
	\vect{S} = \frac{1}{2}\left(\vect{\nabla}\vect{u} + \left(\vect{\nabla}\vect{u}\right)^T  \right) - \vect{I}\left(\frac{1}{3} \vect{\nabla}\cdot\vect{u}\right) \, ,
	\end{equation}
	where $ \vect{I} $ is the identity matrix.
	The pressure is computed by the isothermal ideal gas law, $ p = c_s^2 \rho \, ,$
	where $ c_s $ is the speed of sound. With a constant speed of sound (for the isothermal case) and a constant kinematic viscosity, 
	the momentum equation \rev{(Eq.~\eqref{eq:mom})} can be re-written to:
	\begin{equation}
	\label{eq:momentum}
	\Diff{\vect{u}}{t} = -c_s^2 \vect{\nabla} \left(\ln \rho \right) + \nu \left(\vect{\nabla}^2 \vect{u} + \frac{1}{3}\vect{\nabla}\left(\vect{\nabla} \cdot \vect{u} \right) + 2\vect{S} \cdot \vect{\nabla} \left(\ln \rho \right) \right) \, ,
	\end{equation}		
	which is the form solved in the computations performed in this study.
	
	\subsection{Numerical methods}
	\begin{figure}
		\centering
		\subfigure[IBM]{{\label{subfig:IBM}}\includegraphics[width=0.485\linewidth]{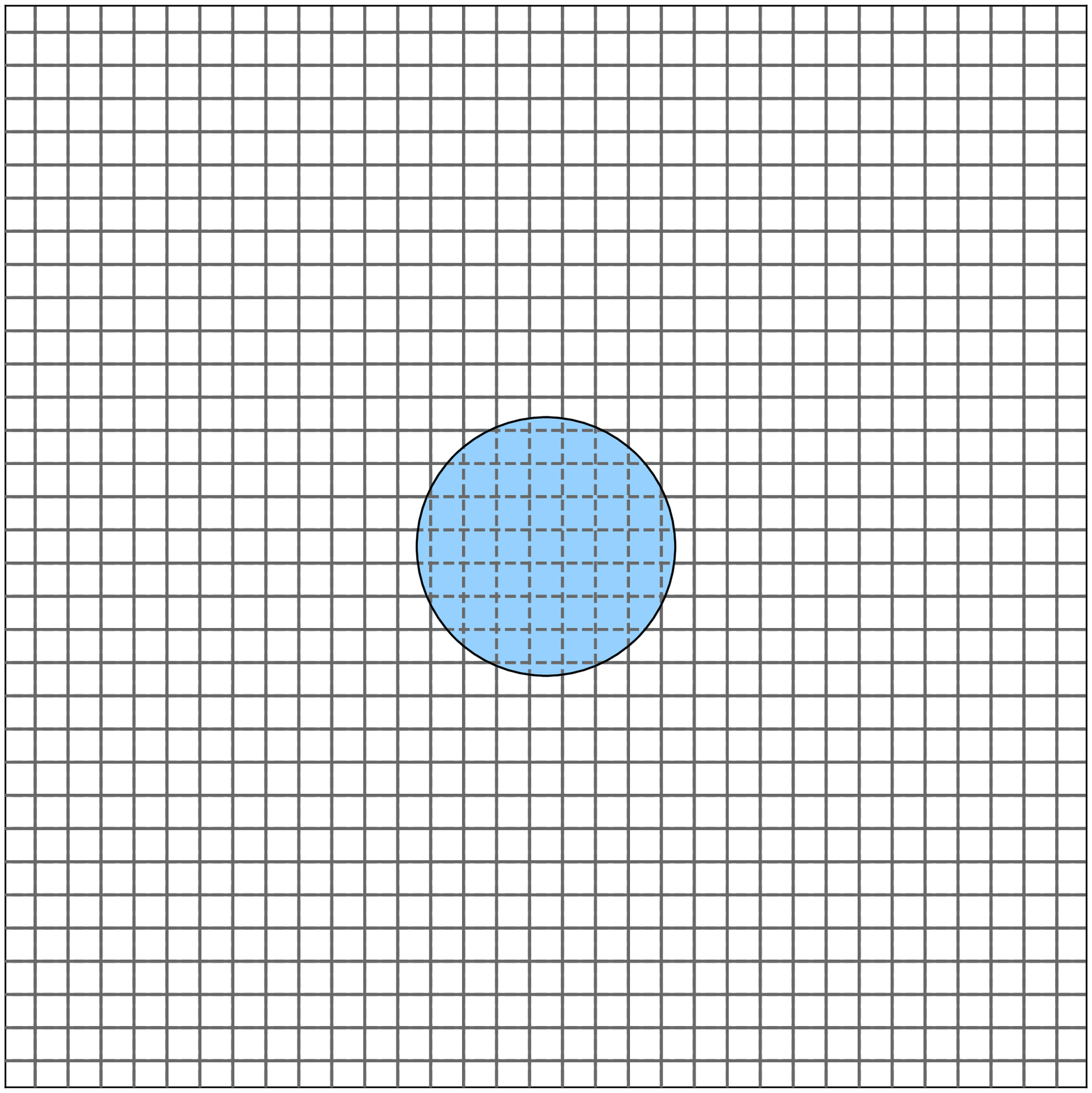}}\quad
		\subfigure[Overset grid]{{\label{subfig:Ogrid}}\includegraphics[width=0.485\linewidth]{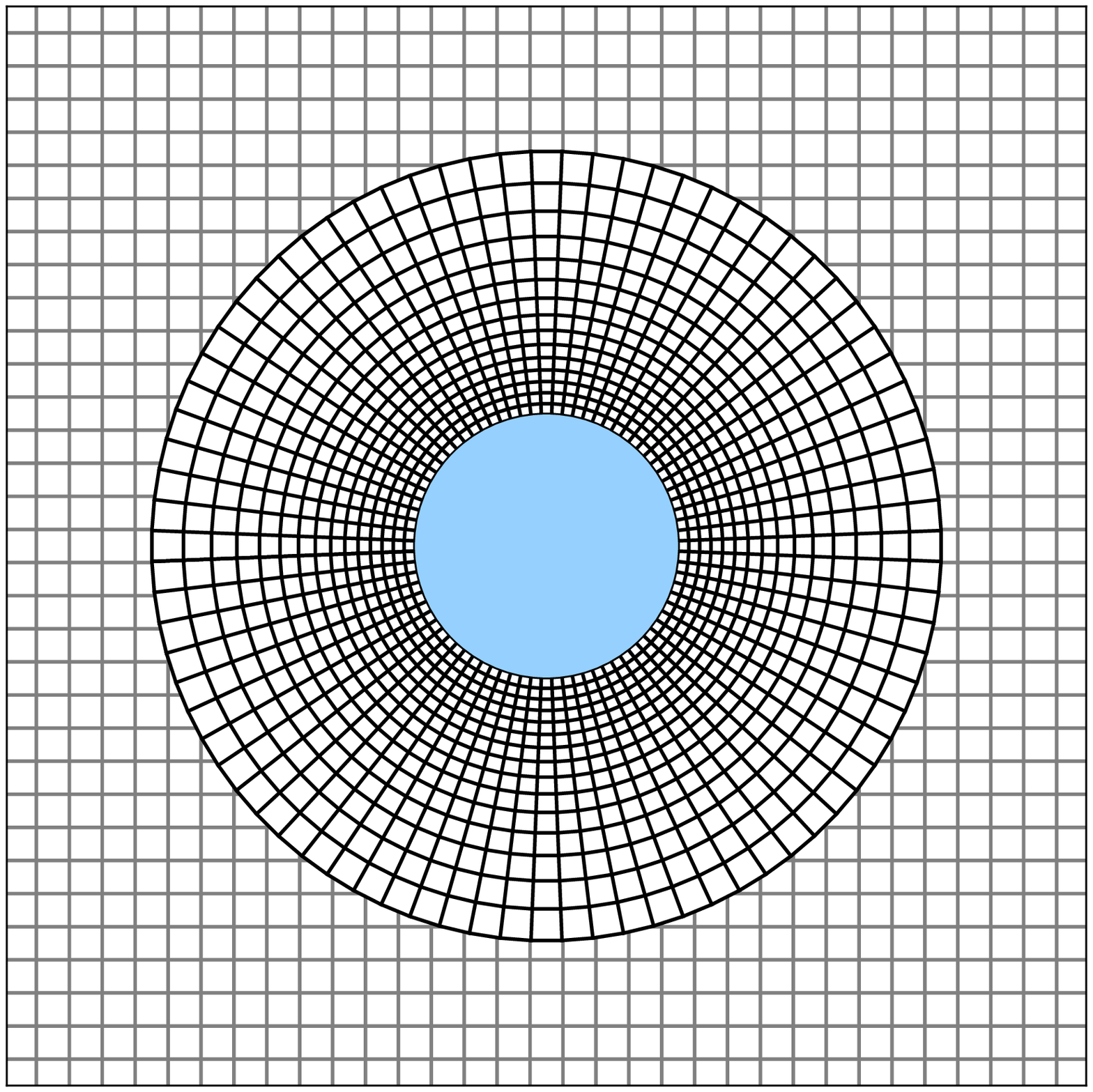}}
		\caption{Solid object representation on a uniform Cartesian grid, by either an immersed boundary method or an overset grid approach. Principle illustration of mesh in each of the cases.}
		\label{fig:solid_rep}
	\end{figure}
	The governing equations (Eqs.~\eqref{eq:continuity} and \eqref{eq:momentum}) are discretized with sixth-order
	finite-differences in space and a third-order memory efficient
	Runge-Kutta scheme in time \citep{williamson1980low}.
	
	Many different types of domain and enforcements of boundary conditions are available in the Pencil Code. For simplicity we consider a domain with a uniform mean flow, using Navier-Stokes characteristic boundary conditions (NSCBC) on both the inlet and the outlet of the flow domain, and periodic boundary conditions in all other directions. The NSCBC is a formulation developed by \cite{Poinsot1992} that makes use of one-dimensional characteristic wave relations to allow acoustic waves to pass through the boundaries.
	
	We place an object in the flow domain by representing it with one of the two available methods in the Pencil Code. The grids used in each of the methods are quite different. An illustration of the grids for IBM and overset grid representation of a circular solid can be seen in Fig.~\ref{fig:solid_rep}. For both methods of solid body representation, we use boundary conditions of no-slip and impermeability for velocity components, and zero gradient for density in the direction normal to the surface, on the solid's surface. The latter condition can be derived from the ideal gas law and the boundary layer approximation \rev{for pressure normal to the boundary ($ \partial p / \partial r = 0 $ \citep{white})}.  In the remainder of this section, details of how the different boundary representations are implemented in the Pencil Code are given.
	
	\subsubsection{Ghost-zone immersed boundary method \label{subsec:IBM}}

	\begin{figure}
		\begin{center}
	\subfigure[Ghost-zone immersed boundary method]{	\includegraphics[width=0.835\linewidth]{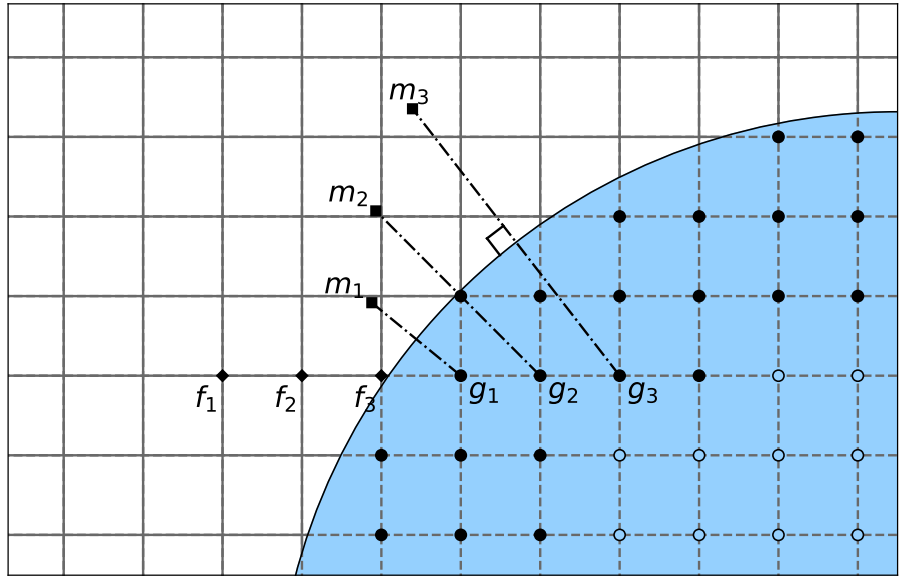}{\label{subfig:IBM_zone}}} \quad
	
	\subfigure[Mirror-point close to surface]{%
	\includegraphics[width=0.4\linewidth]{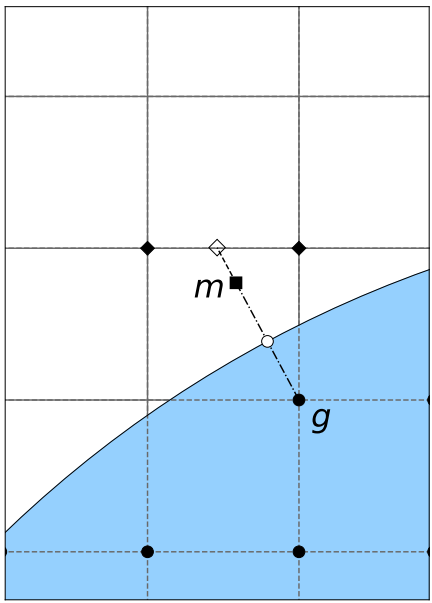}%
	\label{subfig:IBM_mirr_cl}} \quad
	\subfigure[Fluid-point close to surface]{%
	\includegraphics[width=0.4\linewidth]{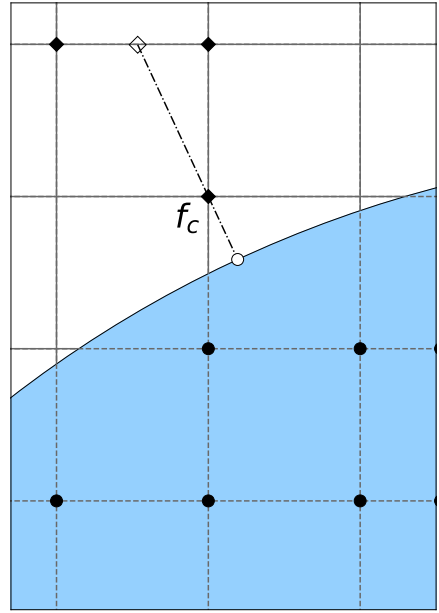}%
	{\label{subfig:IBM_fp_cl}}}
		\end{center}
	\caption{Immersed boundary method. (a): A \rev{zone of ghost-points} ($g_i$; $\CIRCLE$) is used when computing fluid-points ($ f_i $; \sqdiamond). Ghost points are set from corresponding mirror/image-points in the flow domain ($ m_i $; $ \blacksquare $) found along lines orthogonal to the solid surface. The mirror-points are interpolated from surrounding fluid-points. (b): If a mirror-point is too close to the surface to be surrounded only by fluid-points, the values at the points where the orthogonal line intercepts the surface \rev{(}$ \Circle $\rev{)} and the first grid line ($ \Diamond $) are used in interpolation. (c): A fluid-point (\rev{$ f_c $}; \sqdiamond) very close to the surface is set by direct interpolation along the surface normal points by using values at the interception with the surface \rev{(}$ \Circle $\rev{)} and the closest grid line $ (\Diamond) $. The interpolation point at the intersection with the closest grid line in (b) and (c) is set by interpolation from the closest fluid-points along the grid line ($ \sqdiamond $).}
		\label{fig:IBM}
	\end{figure}	
	
	In the illustration of a circular object in a flow domain represented by an IBM (Fig.~\ref{subfig:IBM}) the intersections of solid grid lines represent fluid-points, where the governing equations are solved, while the intersection points of dashed grid lines are grid points inside an immersed solid (solid-points). At the solid-points the governing equations are not solved. Rather, some points are used as ghost-points for the fluid solver and some are unused points. As mentioned, the boundary conditions of a solid object may be imposed directly on the flow variables by the ghost-point immersed boundary method. The IBM in the Pencil Code is such a ghost-point method. An uncommon feature of the IBM implementation in the Pencil Code is that a \rev{several points} deep ghost-zone is used, rather than a single \rev{strip of} ghost-point inside the solid object. This ensures that the \rev{sixth-order finite-difference stencils can be used} without any modifications in the vicinity of a solid object. The overhead related to computation of two additional layers of ghost-points is negligible when compared to the computational cost of the fluid solver itself.
	
	As sixth-order central differencing is used, three points on each side of a grid point are necessary to update the solution. This is illustrated in Fig.~\ref{subfig:IBM_zone}, where stencils of fluid-points $ f_1, f_2 $ and $ f_3 $ will include grid points within the solid object to update the horizontal gradients of the velocity components and density. Point $ f_1 $ will need information from $ f_2 $, $ f_3 $ and $ g_1 $, $ f_2 $ will need information from $ f_1 $, $ g_1 $ and $ g_2 $, etc. (in addition to information from fluid-points to the left). The points in the ghost-zone, $ g_1, g_2 $ and $ g_3 $ are set using corresponding mirror-points $ m_1, m_2 $ and $ m_3 $, respectively. With no-slip and impermeability for velocity and zero gradient for density, the relationship between a ghost point $ g $ and mirror-point $ m $ is simply: 
	\begin{align}
		\revII{\vect{u}(g)} & = -\revII{\vect{u}(m)} \, , \\
		\revII{\rho(g)} & = \revII{\rho(m)} \, . 		
	\end{align}

	Note, that a second-order accurate method to set the Neumann boundary condition has recently been implemented, but will not be described here. Details, and testing of different boundary conditions, can be found in \cite{luo2016ghost}.
	 
	In general, the mirror-points do not coincide with grid points or grid lines, and need therefore to be interpolated from surrounding points. This is done by bi-linear Lagrangian interpolation in 2D and tri-linear Lagrangian interpolation in 3D. For mirror-points close to the surface, one or more of the surrounding grid points may be  inside the solid object (see Fig.~\ref{subfig:IBM_mirr_cl}). Rather than using four surrounding fluid-points, these mirror-points are set by interpolation using the boundary intercept point (that is, the point where the surface normal through the ghost point intercepts the solid's boundary) and the point where the surface normal intercepts the first grid line outside the solid. 
	\rev{Data is first interpolated to the interception point of the surface normal with the first grid line from neighboring fluid-points ($ \sqdiamond $ in Fig.~\ref{subfig:IBM_mirr_cl}) by linear Lagrangian interpolation. The velocity components normal to the surface is expected to scale as $ \Delta r^2 $ when approaching the surface, where $ \Delta r $ is the distance from the boundary. To interpolate velocity in a mirror-point near the surface the velocity is decomposed into cylindrical components, and the radial component is computed by:
	\begin{equation}
	\label{eq:interpspec}
	u_{r,m} =  u_{r,GI} \left(\frac{\Delta r_m}{\Delta r_{GI}}\right)^2 \, ,
	\end{equation}
	where $ u_{r,GI} $ is the radial velocity at the grid line interception point, and $ \Delta r_m $ and $ \Delta r_{GI} $ are the distances from the mirror-point and grid line interception point to the boundary interception point, respectively. Remaining velocity components are obtained by linear interpolation. No special handling is used for density.}
		
	
	Special handling is used for fluid-points very close to the surface of an object. This is done to avoid spurious errors due to de-localization dependencies in the finite-difference stencils, a detrimental effect that occurs when flow variables quite far from a grid point is indirectly used in the update of said grid point. To see this, consider in Fig.~\ref{subfig:IBM_zone} that the horizontal velocity component of grid points surrounding $ m_3 $ will affect $ f_3 $, since $ g_3 $ is set by $ m_3 $. Rather than computing flow variables in a fluid-point close to the surface in the usual way, by using the finite-difference stencils, they are set directly by interpolation, as seen in Fig.~\ref{subfig:IBM_fp_cl}. \rev{The interpolation procedure for these grid points is the same as for mirror-points very close to the boundary, as described above}. Note that this type of special handling is only possible for variables with a Dirichlet boundary condition on the surface.
	
	An alternative to setting mirror-point positions using surface normals is to use mirror-points along grid lines. This simplifies interpolation (making all interpolation one-dimensional, along grid lines) and has proven promising in reducing the errors due to de-localization dependencies. In such an approach, a ghost point can have several values for each flow variable that has a Dirichlet boundary condition, one used in horizontal and one in vertical finite-difference stencils (as would be the case for $ g_1, g_2 $ and $ g_3 $ in Fig.~\ref{subfig:IBM_zone}). Flow variables with Neumann boundary conditions are set from mirror-points along surface normals as in the method described above. In this study, we will stick to the more mature method of using mirror-points from surface normals. Details on the alternative grid-line ghost-zone IBM can be found in \rev{\cite{aarnes2018}}.

	\subsubsection{Local-time restricted overset grids {\label{subsec:ogrid}}}
	Unlike the representation of solid bodies with most methods (IBM, body-fitted structured or unstructured grids), codes using overset grids require splitting of the flow solver, as one solver is needed for each grid. We limit this study to a single grid on top of a background grid, for a more general discussion see \cite{Chesshire1990} or \cite{Meakin1995}. \revII{Yin-yang grids are not considered (although a yin-yang grid implementation exists in the Pencil Code).}
		
	For a flow with a solid object represented using a body-confined grid over a Cartesian background grid, the governing equations are, in principle, solved for two different flow domains, one with and one without a solid object present in the flow. At least one boundary in each domain is set by interpolating flow variables from another grid, and in this way the presence of the solid affects the flow on all grids. Admittedly, this makes overset grids somewhat more unwieldy than IBMs. 
	
	To consider this more systematically, let us regard a fluid time step as split into four parts: (1) solution of the governing equations on the background grid, (2) communication of data from the background grid to the body-fitted grid, (3) solution of the governing equations on the body-fitted grid, (4) communication of data from the body-fitted grid to the background grid. The solution step on the body-fitted grid requires implementation of a Navier-Stokes solver applicable to the type of grid that is used to resolve the bluff body's boundary. For our case of a cylinder in a cross flow, a Navier-Stokes solver applicable to cylindrical coordinates is necessary.
	
	In the illustration of the mesh used in the \rev{(cylindrical)} overset grid method (Fig.~\ref{subfig:Ogrid}) there are no grid points inside the circular object. Strictly speaking, the background Cartesian grid is present in the entire domain (also inside the limits of the curvilinear mesh and inside the solid object), but \rev{only} few points inside the curvilinear mesh are used, and not a single Cartesian grid point inside the solid is (or should ever be) used. 
	
\begin{figure}
		\centering
		\subfigure[Interpolation: Cartesian to curvilinear grid]{%
			\includegraphics[width=0.65\linewidth]{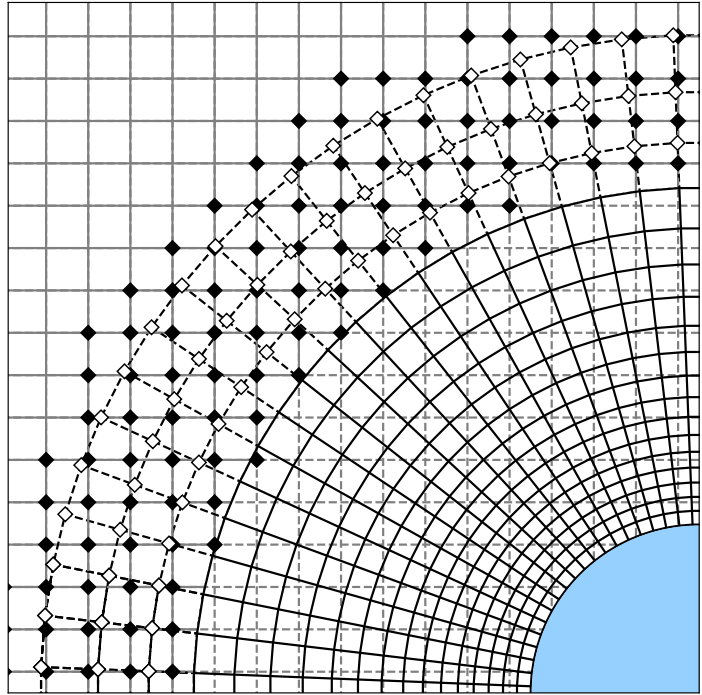}%
			\label{subfig:cart_to_curv}}
		\subfigure[Interpolation: Curvilinear to Cartesian grid]{%
			\includegraphics[width=0.65\linewidth]{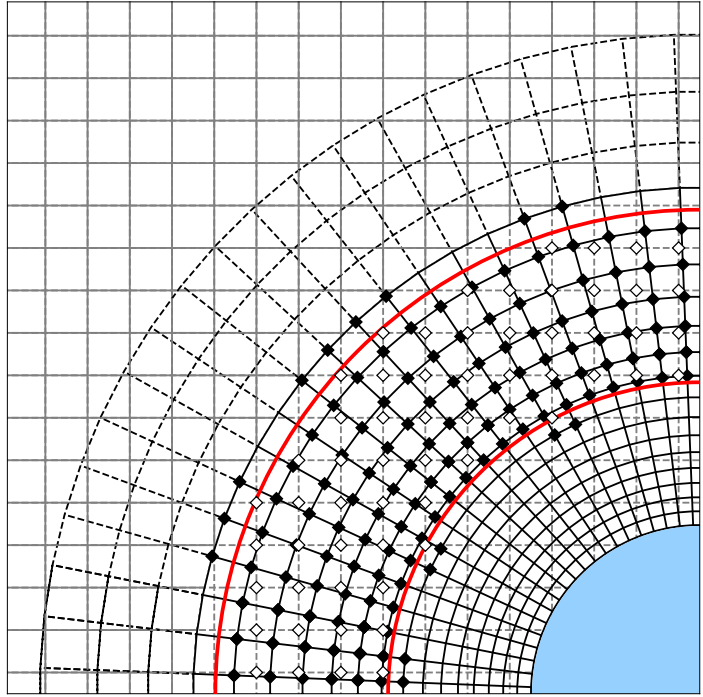}%
			\label{subfig:curv_to_cart}}
		\caption{Overset grid method. Interpolation between grids, from donor-points (\sqdiamond) to fringe-points ($ \Diamond $). The outer points on the curvilinear grid are by default set as fringe-points, while on the Cartesian grid a zone of fringe-points \rev{are identified} during pre-processing. Cartesian grid points closer to the solid than the inner diameter setting the fringe-point zone are hole-points. \rev{Intersections of solid grid lines represent regular fluid-points where finite-difference stencils are used to update the flow variables. At dashed cylindrical grid lines all intersections are fringe-points. At dashed Cartesian grid lines, intersections may be regular fluid-points, fringe-points or hole-points. Fringe-points are identified in-between the inner and outer interpolation zone radius (red circular lines) set according to selected interpolation scheme. Cartesian grid points closer to the solid than the inner interpolation radius are hole-points.}}
		\label{fig:ogrid_comm}
	\end{figure}

	The points of the background grid that are in use inside the curvilinear mesh are mostly fringe-points. Fringe-points are fluid-points set by interpolation from nearby donor-points on the overlapping grid, rather than computed using discretization of the governing equations. Donor-points are computed in the same way as an ordinary fluid-point, unless the interpolation is implicit, meaning that a fringe-point may be used as a donor-point \citep{Chesshire1990}, which is not the case here. A third class of points found on overset grids are hole-points. These are unused grid points, typically found outside of the fringe of a grid (like Cartesian grid points far inside the curvilinear grid). Fringe-, donor-, and hole-points on an overset grid generated to represent a circular object are seen in Fig.~\ref{fig:ogrid_comm}.

	During the first stage of the inter-grid communication, data is sent to the outermost grid points of the body-fitted grid (Fig.~\ref{subfig:cart_to_curv}). At the second stage, data is sent back to Cartesian fringe-points (Fig.~\ref{subfig:curv_to_cart}). 
	The fringe-points on the Cartesian grid are identified during pre-processing, where points \rev{within a distance from the solid, illustrated by red circular curves in Fig.~\ref{subfig:curv_to_cart}, are selected}. 
	For moving objects, fringe-point locations on the background grid must be re-calculated every time step. A \rev{several layers thick} zone of fringe-points is used on both grids to enable the use of the sixth-order centered stencils at the outer edges of each grid, equivalent to the use of a ghost-zone for the IBM. 
	
	In the illustration in Fig.~\ref{fig:ogrid_comm} each fringe-point is surrounded by four donor-points, as necessary in bi-linear Lagrangian interpolation. If higher order interpolation is desired the amount of donor points for each fringe points must be increased accordingly. Such an increase is straightforward for overset grids, unlike for immersed boundary methods where more intricate interpolation stencils are needed for high-order interpolation to avoid using grid points that are inside the bluff body. Note, however, that a straightforward extension from second to third order interpolation (or higher) does not guarantee a better solution. This is due to possible overshoots in the interpolation polynomials. High-order Lagrangian interpolation and quadratic splines are implemented for overset grid interpolation in the Pencil Code. For simplicity, we will restrict ourselves to bi-linear Lagrangian interpolation here.
	
	At the solid-fluid interface, we use summation-by-parts (SBP) finite-difference operators to enhance stability of the solution. This means modifying the finite-difference stencils in the nine grid points closest to the surface (including the surface point) along each radial grid line, to asymmetric stencils (one-sided at the surface). The order of accuracy for the SBP-operators are third-order for a sixth-order finite-difference method. Details on these operators can be found in \cite{Strand1994} (first derivatives) and \cite{Mattsson2004} (second derivatives).
	
	A peculiar feature of the overset grid implementation in the Pencil Code is how the restrictions on the time step are handled. The advective and diffusive time step restrictions are $ \Delta t \leq C_{\nu} \Delta \chi_{min}^2 / \nu $ and $ \Delta t \leq C_u \Delta \chi_{min} / \left( \left| \vect{u} \right| + c_s\right)$, respectively, where $ \Delta t $ is the time step, $ \Delta \chi_{min} $ the smallest grid spacing in any direction, and $ C_\nu $ and $ C_u $ are the diffusive and advective Courant numbers, respectively. For a weakly compressible flow we typically require a very short time step, increasingly so if grid stretching is used in order to have a fine grid in the vicinity of the solid object. However, when overset grids are used these restrictions are no longer global restrictions on the time step, but local. Hence, by performing several time steps on the body-fitted grid for each time step on the background grid, the efficiency of the code may be greatly improved. In particular, this allows for a very fine resolution close to the surface (on the body-fitted grid) without using small time steps for flow far from the surface (on the background grid).

	\rev{For all the overset grid computations in the present study the diameter of the cylindrical grid is three times that of the solid cylinder it is fitted to. For a consideration of the extent of the domain covered by the body-fitted grid the reader is referred to \citet{aarnes2018ogrid}.}
	\subsubsection{A note on dissipation}
	The centered finite-difference schemes used for discretization of the governing equations are non-dissipative. This can cause problems due to the potential growth of high-frequency modes, leading to	numerical instability. 
	
	To some extent, the summation-by-parts boundary conditions
	suppress such instabilities that are due to boundary conditions when overset grids are used, but these boundary stencils are not sufficient to suppress all oscillations in the solution when grid stretching is used on the curvilinear grid. Such oscillations are most prominent in the
	density field. The detrimental effect of the high-frequency modes increase\rev{s}
	as the grid spacing decreases, and may lead to diverging solutions as
	the grid is refined. To suppress the high-frequency modes, a
	high-order low-pass filter is used on the curvilinear part of the
	overset grid. The filter is a 10th order Pad\'{e} filter, with
	boundary stencils of 8th and 6th order. On the interior of the domain, the filter is given by: 
	
	\begin{equation}
	\alpha_f \hat{\phi}_{i-1} + \hat{\phi}_{i} + \alpha_f \hat{\phi}_{i+1} =
	\sum_{n=0}^{N} \frac{\alpha_n}{2} ({\phi}_{i+n} + {\phi}_{i-n}) \, ,
	\end{equation}
	where $ \hat{\phi}_k $ and $ {\phi}_k $ are components $ k $ of the filtered and unfiltered solution vectors, respectively, $ \alpha_f $ is a free parameter ($\left| \alpha_f \right| \leq 0.5$) and $ \alpha_n $ are fixed parameters dependent only on $ \alpha_f $ \citep[details in][]{Visbal1999}. Boundary stencils can be found in \cite{Gaitonde2000}.
	The Pad\'{e} filter is implicit, and requires us to solve a tri-diagonal linear system at each grid point, in the radial direction, and a cyclic tridiagonal system in the direction tangential to the surface. The free parameter $ \alpha_f$ is set to 0.1. With such a small value for $ \alpha_f $, filtering the solution once per Cartesian time step is found sufficient to get a stable and accurate solution. 
	
	\rev{Alternatively, a sixth-order hyper-diffusion operator, which is already implemented in the Pencil-Code \citep[see e.g.][]{haugen2006hydrodynamic}, could  have been used to filter the solution. The benefits of this approach is that the hyper-diffusion operator is explicit and fast, and does not require extra communication between processors. It is, however, expected to be less sharp than the 10th order Pad\'{e} filter, as Pad\'{e} filters are known to outperform explicit filtering schemes \citep{Visbal1999,visbal2002use}.}
	
	When IBM is used rather than overset grids, some dissipation can be turned on by using fifth-order upwinding for the advection operators of the density rather than central-difference stencils \citep[details in][]{Dobler2006}. The problem of oscillations in the density field is, however, much less prominent when a uniform Cartesian mesh is used, so while Pad\'{e} filtering is on by default when overset grids are used, a dissipative solution by upwinding is optional for other simulations.
	
	\section{Simple geometry {\label{sec:benchmarking}}}
	In previous studies, the order of accuracy of the solid object representations described above has been assessed for steady flow computation. Using a slightly modified handling of Neumann boundary conditions, \cite{luo2016ghost} showed that, regardless of boundary condition, the IBM implementation in the Pencil Code is second-order accurate in the vicinity of a resolved circular boundary. For the same geometry, using overset grids, \cite{aarnes2018ogrid} showed that the order of accuracy \revII{in the vicinity of the solid} differed for different \revII{flow variables.} 
	The radial velocity component was computed with order of accuracy between third and fifth-order, while the accuracy of the tangential velocity \revII{and density} was between second and third-order, when second-order Lagrangian interpolation \revII{was used for communication between grids}.
	Both the mentioned studies also demonstrated that characteristic flow parameters, like drag, lift and shedding frequency (for unsteady flow), could be reproduced to good agreement with previous studies, with the respective boundary representation in use. 
	
\begin{figure}
		\centering
			\subfigure[{$ Re=20 $}]{%
				\includegraphics[width=0.325\linewidth]{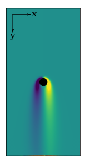}%
				\label{subfig:visuRe20}}
			\subfigure[{$ Re=100 $}]{%
				\includegraphics[width=0.325\linewidth]{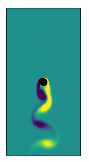}%
				\label{subfig:visuRe100}}
			\subfigure[{$ Re=400 $}]{%
				\includegraphics[width=0.325\linewidth]{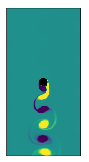}%
				\label{subfig:visuRe400}}
		\caption{Flow visualization. Contours of instantaneous vorticity $ \omega_z = \left[\nabla \times \vect{u}\right]_z $ (normal to the view plane) plotted for three different Reynolds number\rev{s}. Inflow at the top of plane.}
		\label{fig:visualization}
	\end{figure}

	We will not repeat an assessment of accuracy for steady flow computations here. Rather, we investigate the boundary representations by a direct comparison for the case of flow past a circular cylinder in different shedding regimes. Two-dimensional flow past a circular cylinder in the vortex shedding regime is a classical benchmarking case for fluid dynamic simulations with a solid object present in the flow domain. We simulate such a flow with Reynolds number 100, a Reynolds number where the von K\'{a}rm\'{a}n vortex street can be observed in the cylinder's wake. The Reynolds number is defined as $ Re = U_\infty D/\nu $, where $ U_\infty $ is the \rev{incoming} flow velocity and $ D $ is the diameter of the cylinder obstructing the flow. In addition, we test each boundary representation for a steady flow ($ Re=20 $) and an unsteady flow with more chaotic tendencies ($ Re=400 $). Note that three-dimensional effects in the latter flow are suppressed as we restrict ourselves to a two-dimensional domain. A rectangular domain with domain size $ L_x \times L_y = 10D \times 20D $ is used, with a body-fitted grid with diameter $ 3D $ used in the overset grid simulations. For overset grids, grid stretching is used in the radial direction to obtain approximately quadratic cells at the cylinder surface and in the region where \rev{the} interpolation between the grids is performed. In this region the cells of both grids are similar in size. The inflow Mach number ($ Ma = U_\infty/c_s $) is set to 0.1, and the Reynolds number is varied by adjusting the value of the kinematic viscosity. The vorticity component normal to the $ xy$-plane  for the three different Reynolds number flows \rev{with mean flow in the $ y $-direction} can be seen in Fig.~\ref{fig:visualization}.
	
\begin{figure}
	\centering
	\includegraphics[width=0.7\linewidth]{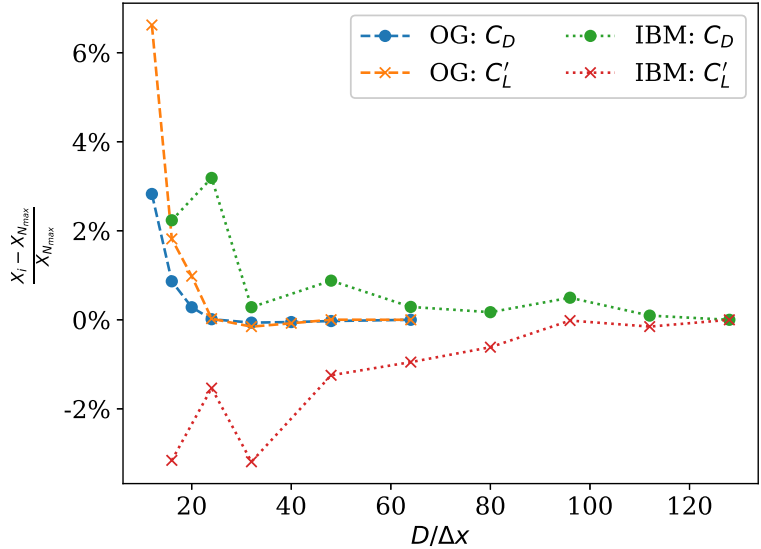}
	\caption{Normalized values for mean drag coefficient $(C_D) $ and root-mean-square lift coefficient $(C_L')$ for flow with $ Re=100 $ computed at grids with varying coarseness. The cylinder in the flow is represented either by the immersed boundary method (IBM) or with overset grids (OG).}		
	\label{fig:gridrefRe100}
\end{figure}	
	
	Consider Fig.~\ref{fig:gridrefRe100}, depicting normalized deviations of mean drag coefficient and root-mean-square lift coefficient computed at different resolutions with the two methods for $ Re=100 $. For both cases, the results from the finest grid is used for normalization. Strouhal number (dimensionless shedding frequency) is not included in the figure, since it is barely affected by the grid spacing and is therefore not a good measure of grid independence.  
	The number of grid points per diameter, on the Cartesian grid, is given on the horizontal axis. Note that this might be somewhat misleading, as the overset grid uses two grids to cover the flow domain. For the flow domain and grid sizes used in this study, a\rev{n} overset grid simulation uses 10\% more grid points than a corresponding IBM simulation when $ D/\Delta x $ of the two simulations \rev{is} the same. No matter this difference, it is clear that the overset grid method greatly outperforms the IBM with respect to the necessary grid required to reach grid independency under the conditions of these simulations. Using a background grid with $D/\Delta x \geq 24 $ the deviation from results on a $D/\Delta x = 64 $ grid is less than 0.1\rev{6}\% for the overset grid. The results from the IBM calculations converge much slower. To reach a comparable level of grid independence to that of the overset grid with $ D/\Delta x = 24 $, a grid using IBM requires $ D/\Delta x\geq 112 $. Such a fine grid yields a deviation of less than $ 0.1\% $ in drag and $ 0.1\rev{6}\% $ in lift, from the results at the finest grid level ($ D/\Delta x = 128 $). If these grids ($ D/\Delta x = 24 $ for overset grids, $ D/\Delta x = 112 $ for IBM) are \rev{deemed} sufficiently accurate resolutions for grid independent solutions for the different solid object representations, the IBM requires 18.1 times as many grid points as the overset grid method on the two-dimensional domain used in these simulations. In these simulations, the advective restriction is more strict than the viscous restriction, hence, the time step is proportional to the grid spacing. This means that there is a factor 4.7 difference in time step between the $ D/\Delta x = 24 $ and $ D/\Delta x = 112 $, that has an additional large impact on the (in)efficiency of the IBM as compared to the overset grids.
	
	For practical application, it may perhaps be excessive to require $ \approx 0.1 \% $ deviation for results to be deemed grid independent. With a resolution $D/\Delta x \geq 64 $ in the simulations performed with IBM, there is less than 0.5\% deviation in drag and less than 1\% deviation in lift. Choosing such a resolution will, in many cases, be an acceptable trade-off between accuracy and efficiency. This reduces the difference between the overset grid and IBM somewhat, although it warrants the use of a somewhat coarser grid for overset grid computations as well. Note that in computing drag and lift forces on the cylinder represented by the IBM, so-called force-points are used. The force-points are distributed uniformly around the cylinder, and viscous and pressure forces are approximated at these points, using data from surrounding grid points. Some of the oscillations in the computed mean drag and root-mean-square lift coefficients seen in Fig.~\ref{fig:gridrefRe100} may be due to a change in the position of force-points when the grid is refined. 
		
	\rev{A relevant consideration when the different costs associated with overset grids and IBM are compared is the computational cost of interpolation in the two different methods. With equally spaced Cartesian grids, more fringe-points are interpolated with overset grids than mirror-points interpolated with the ghost-point IBM method. This is due to overset grids having two zones of interpolation (one for interpolation from Cartesian to cylindrical, and one for interpolation back to Cartesian), a larger circumference of the interpolation regions (interpolation farther from the cylinder) and the need for a deeper ghost-zone on the Cartesian grid in overset grids since no special handling for fluid-points close to the fringe-point region is used. For the $ D/\Delta x = 24 $ grid, the total number of fringe-points for the overset grids is approximately 1600 ($ < 1.3 $\% of total number of grid points). Note that the fraction of fringe-points to grid points decreases as the grid spacing decreases ($<0.65\%$ grid points are fringe-points when $ D/\Delta x = 48 $). The number of mirror-points in an IBM simulation with the same grid spacing is approximately 200. A fringe-point is updated only once every Runge-Kutta time step, while mirror-points are updated every sub-time step. Hence, approximately 2.5 times as much interpolation is performed when the solid is represented by overset grids rather than by IBM, \revII{if the same grid spacing is used in the different solid object representations. As much finer grids are required with IBM than with overset grids, the advantage of a smaller interpolation cost with IBM is lost.}}
		
	\begin{table}[t]
		\centering
		\caption{Comparison with data sets from previous studies for $ Re=100 $. Asterisk denotes scaled values of $ C_L $. The non-rectangular grids are marked as circular inlet/C-type ($\Leftcircle$) or circular/O-grid ($\Circle$). Domains in which the cylinder is not centered have both upstream and downstream lengths given.}
		\label{tab:benchmarking}
		\begin{tabular}{l c *{3}{l}}
			\hline
			& $\left[(L_{x_u} + L_{x_d}) \times L_y\right]/D^2 $ & $C_D$ & $C_L'$ & $St$\Tstrut\Bstrut \\
			\hline
			\cite{Kim2001}         
			& $70\times 100$ 			& 1.33   &   $0.22^{(*)}$	&  0.165\Tstrut   \\
			\cite{Pan2006}       
			&  $60\times 60$				& 1.32   &   $0.226^{(*)}$ 	&  0.16    \\
			\cite{Haugen2010}   
			&  $70 \times 35$			& 1.328  &   $\quad -$ 	    &  0.166   \\
			\cite{Park1998}        
			&  $(50+20) \times 100$, $\Leftcircle$ & 1.33   & $0.235^{(*)}$  			&  0.165   \\
			\cite{Shi2004}            	&  $300$  $\Circle$ & 1.318  &   $\quad -$ 	    &  0.164 \\
			\cite{Mittal2005a}	
			&  $100 \times 100$ & 1.322  & 0.226  			&  0.164  \\
			\cite{Stalberg2006}   
			&  $160$,  $\Circle$ & 1.32   & $0.233^{(*)}$	&  $\quad -$   \\
			\cite{Li2009}					& $100 \times 100$  & 1.336  &   $\quad -$ 	    &  0.164   \\	
			\cite{Posdziech2007}  & $(20+50) \times 40$ $\Leftcircle$ &  1.350 & $0.234^{(*)}$  &  0.167  \\
			\cite{Posdziech2007}  & $(4000+50) \times 8000$ $\Leftcircle$ &  1.312 & $0.224^{(*)}$  &  0.163  \\
			\cite{Qu2013}               	& $60 \times 60$ & 1.326  & 0.2191 			&  0.166 \\
			\cite{Qu2013}               	& $200 \times 200$ & 1.310  & 0.2151 			&  0.165 \\
			Present, IBM                   			&  $50 \times 50$		& 1.351  & 0.232  			&  0.166  \\
			Present, overset grid.	&  $50 \times 50$	&	1.3\rev{47} & 0.23\rev{4} & 0.16\rev{6}\Bstrut \\
			\hline 
		\end{tabular}
	\end{table}
	
	To verify that the flow is computed accurately, the resolutions from the discussion above ($ D/\Delta x = 24 $ for overset grids, $ D/\Delta x = 64 $ for IBM) are used in a simulation on a large domain for each of the solid body representations. The domain size is set to $ L_x = L_y = 50D $. The resulting mean drag, root-mean-square lift, and Strouhal frequency are compared to results reported from other studies, in Tab.~\ref{tab:benchmarking}. Domain sizes and types are listed in the table, along with the most relevant flow coefficients. Some of the listed values for root-mean-square lift coefficients are scaled values, as only amplitude of the lift coefficient was reported from these particular studies. A scaling factor of $1/ \sqrt{2} $ has been used (since the lift coefficient is a smooth sinusoidal-like function with zero mean value), and the scaled results are marked with a superscript (*).	The studies in Tab.~\ref{tab:benchmarking} use a wide range of numerical methods to compute the flow, including finite-volume, finite-difference, finite-element, spectral element and lattice-Boltzmann methods. \rev{The top three studies in Tab.~\ref{tab:benchmarking} use immersed boundary methods to represent the solid cylinder, while the remaining studies (present IBM simulations excluded) use body-fitted methods.} Only \cite{Haugen2010} and \cite{Li2009} simulate compressible flows (where the former of these uses the Pencil Code with the IBM described here, but with different domain size and resolution). Table \ref{tab:benchmarking} includes two results from each of the studies by \cite{Posdziech2007} and \cite{Qu2013}, to include results from both comparable domain sizes to the present study, and highly accurate results from very large domains. The present results, both those computed with the IBM and the results from overset grid simulations, agree well with the results found in \rev{the} literature.
	
\begin{figure}
			\subfigure[$ Re=20 $]{%
				\includegraphics[width=0.49\linewidth]{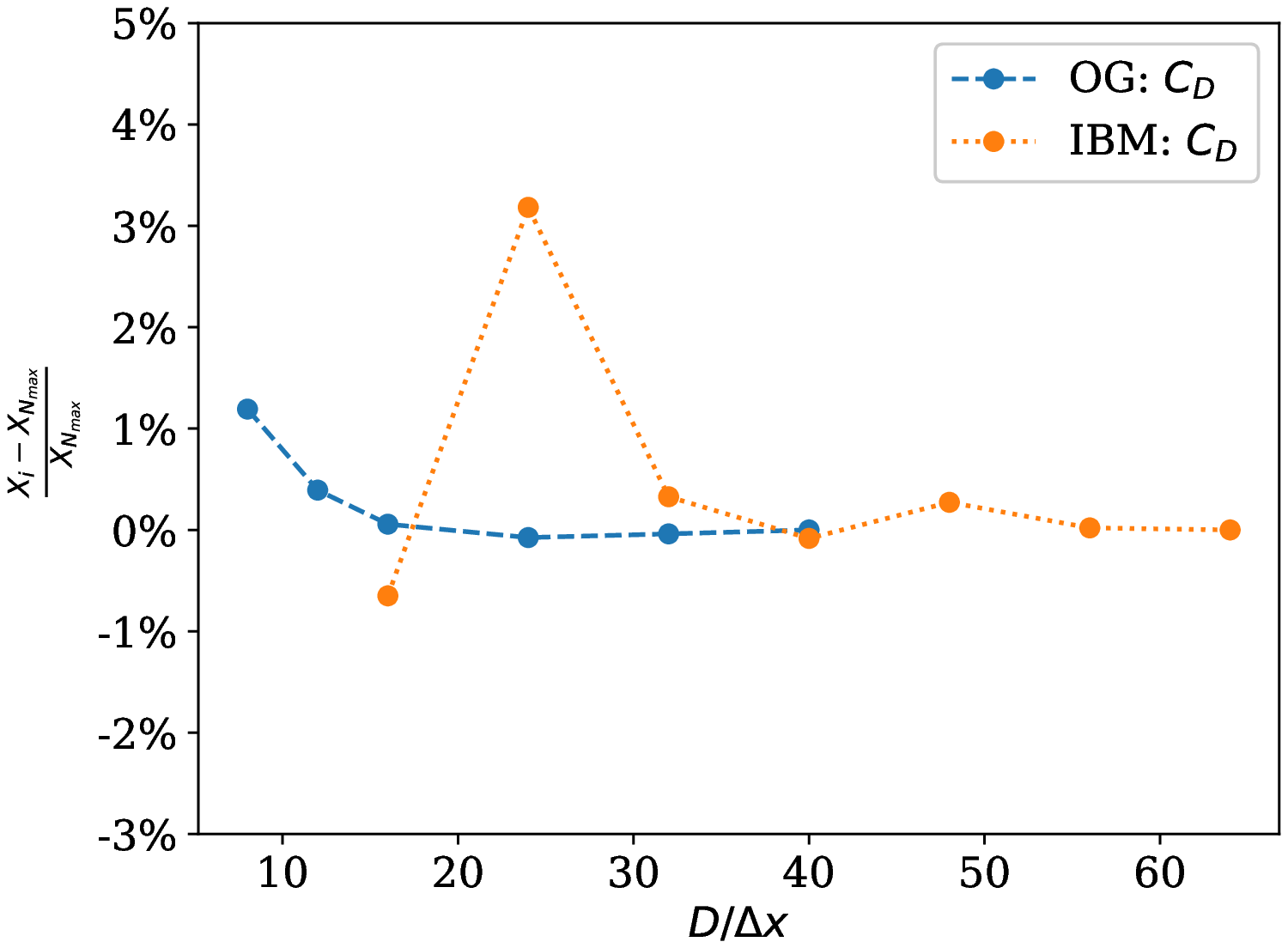}%
				\label{subfig:gref20}}
			\subfigure[$ Re=\rev{400} $]{%
				\includegraphics[width=0.49\linewidth]{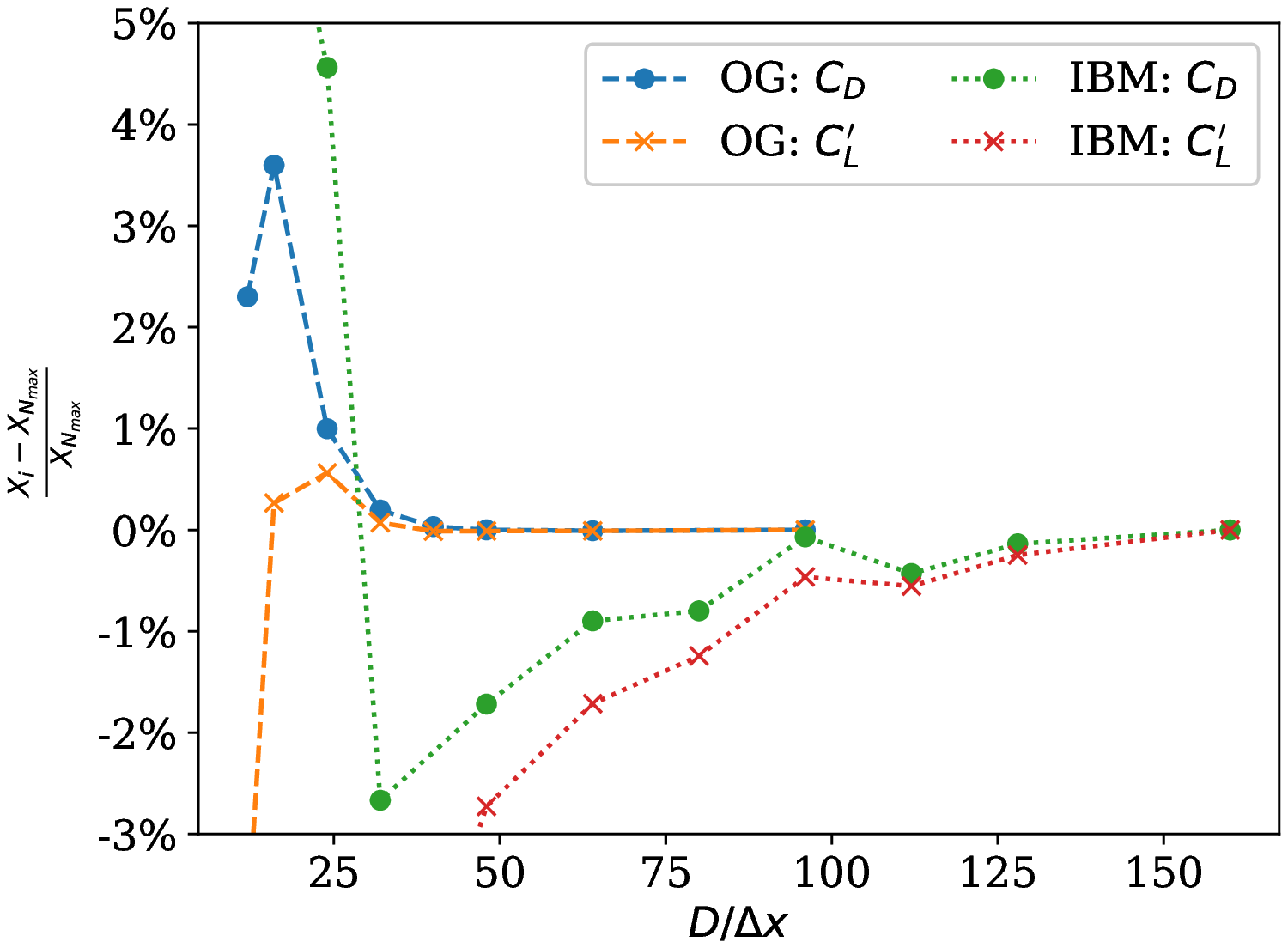}%
				\label{subfig:gref400}}
			\caption{Normalized values for mean drag coefficient $(C_D) $ for flow with $ Re=20 $, mean drag coefficient and root-mean-square lift coefficient $(C_L')$ for flow with $ Re=400 $. Results are computed at grids with varying coarseness. The cylinder in the flow is represented either by the immersed boundary method (IBM) or with overset grids (OG).}		
			\label{fig:gridrefRe20_400}
		\end{figure}

	\rev{Grid independence results for $ Re = 20 $ and $ Re=400 $ are depicted in Fig.~\ref{fig:gridrefRe20_400}. The results are similar to those obtained for $ Re=100 $:
	there is} a much more rapid convergence to grid independent solutions with overset grids than with IBM. With overset grids, a background grid with $ D/\Delta x\geq 32 $ yields less than 0.2\% deviation in the drag and lift coefficients, at $ Re=400 $. This means using overset grids $ N_x \times N_y +N_r \times N_\theta= 320 \times 640 +    64 \times  320$. With IBM, a grid with $ D/\Delta x\geq 128$ is necessary to get comparable grid independence (deviation less than 0.25\% in drag and lift coefficients). That means using a $ 2560 \times 1280 $ grid for this specific domain size, and a factor four larger time step than on the background grid in the overset grid simulation. For the steady flow, the lift coefficient is not defined, and for this reason we include only the drag coefficient in the grid independence comparison. The results in Fig.~\ref{subfig:gref20} show that $ D/\Delta x =12$ is sufficient to obtain drag with less than 0.4\% deviation from the finest grid result for the overset grid computation. With IBM, $ D/\Delta x =32$ is needed to get the deviation down to the same level.
	
	From these tests it is clear that in representing the simple geometry of a circular cylinder with the Pencil Code, the method of overset grids is far superior to the IBM in terms of efficiency and accuracy. Not only is far less grid points required (and, consequently a larger time step allowed) to reach a grid independent solution, there is also far less variation in the solution before grid independence is reached. With the IBM we may have to accept a deviation of, say, 1.0\% in mean drag and root-mean-square lift coefficients from one grid to a finer one. In the case of overset grids, on the other hand, a deviation one order of magnitude smaller than this can be achieved at reasonable computational costs. As mentioned, however, some of the variation seen in the IBM results may be attributed to the way the coefficients themselves, and not the flow, are computed. The positioning of force-points where drag and lift are computed is affected by the choice of resolution, but has no influence on the solution of the flow equations. 
	
	That being said, we should now address the limitations of the overset grid method. In short, the problem of adaptability to different geometries is a major drawback of this method. Even an extension from a 3D cylinder to a sphere would require a completely new grid handling, with updates needed all the way down to the level of finite-differences in the code. For more complex shapes, where an analytic transformation from Cartesian space to the fitted grid coordinates is not available, this will become increasingly difficult, if not impossible. It is in this respect that the full potential of the IBM can be achieved. The simple handling of boundaries and lack of any modification needed in the treatment of the governing equations, make the  ghost-zone IBM ideal for complex geometries of all kinds. How this is done in the Pencil Code is the topic of the remainder of this paper.
	\section{Complex geometries {\label{sec:complex}}}
	
	\begin{figure}[t]
		\centering
		\subfigure[Identifying fluid/solid-points]{%
			\includegraphics[width=0.485\linewidth]{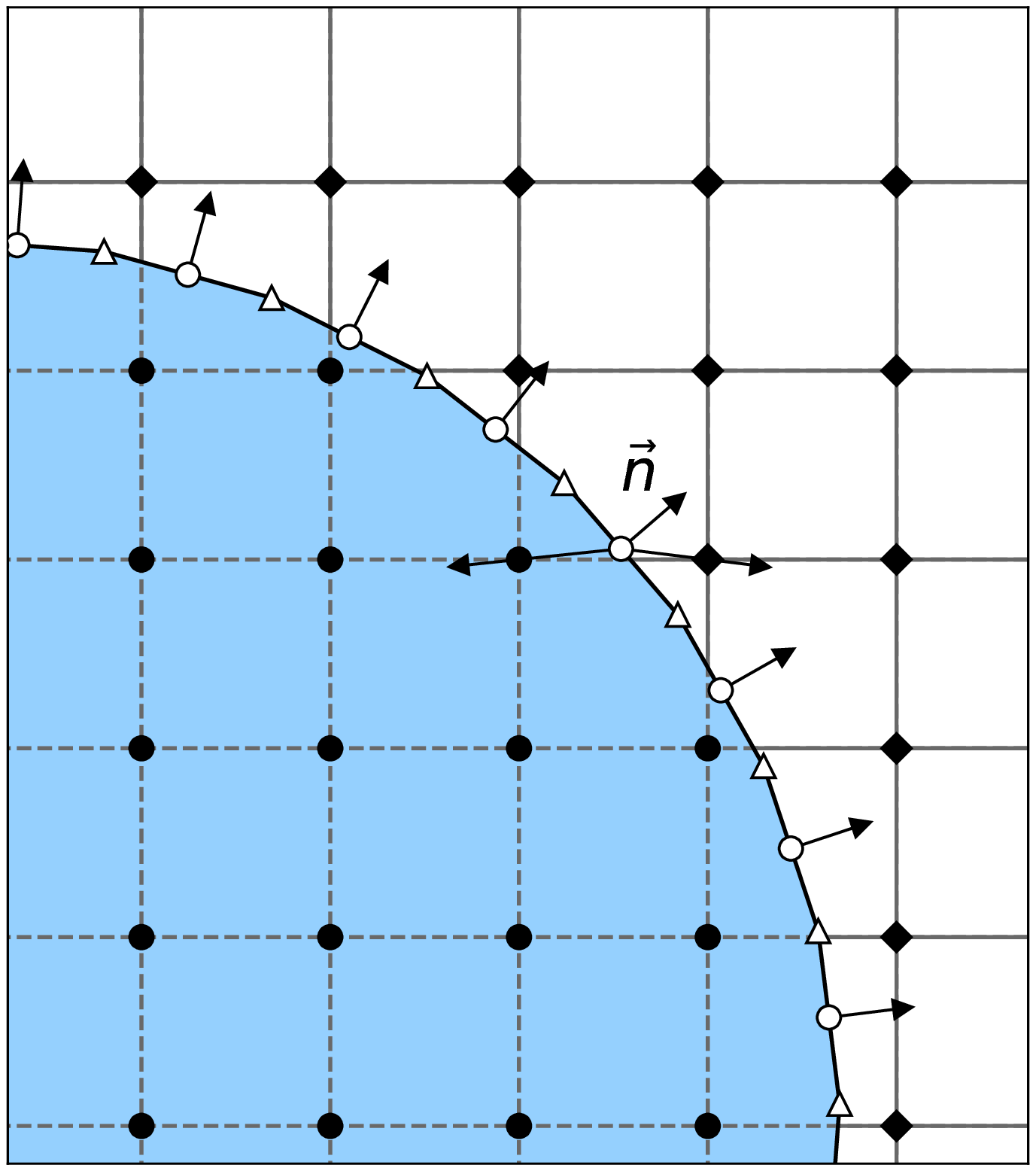}%
			\label{subfig:IBM_complx_points}}\quad
		\subfigure[Mirror-point computation]{%
			\includegraphics[width=0.485\linewidth]{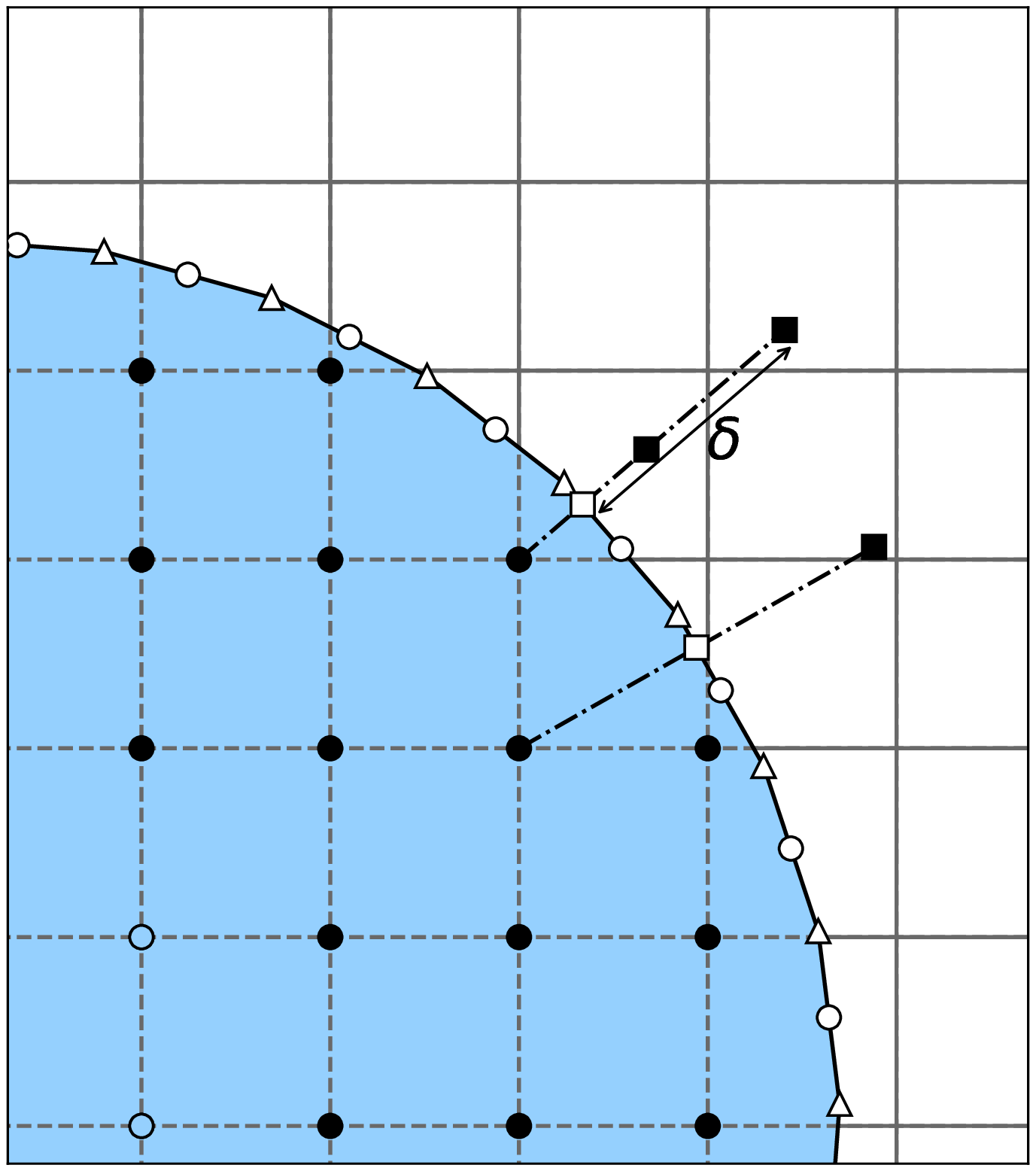}%
			\label{subfig:IBM_complx_mirror}}
		\caption{Immersed boundary method. (a): Identifying grid points as solid-points or fluid-points for a two-dimensional geometry by dot product of line-segment normal and directional vector grid points. (b): Schematic diagram for ghost points and the method to assign mirror/image-points along lines normal to line-segments between vertex points. ($\CIRCLE$) solid-point, (\sqdiamond) fluid-point, ($ \Circle $) centroid of line-segment, ($ \Delta $) vertex of line-segment, \rev{($ \vec{n} $) line-segment's normal vector}, ($ \square $) boundary intersect point, ($ \blacksquare $) mirror-point.}
		\label{fig:IBM_complex}
	\end{figure}
	One of the difficulties in \rev{extension to irregular geometries of}
	a ghost-cell immersed boundary method's 
	lies in how to track the boundaries correctly. To the best of our knowledge, two ways to overcome this exist: the unstructured triangle surface mesh \citep{gilmanov2003,mittal2008versatile,nagendra2014new} and the combination with level-set signed distance function\rev{s} \citep{liu2014efficient,uddin2014cartesian}. 
	The first method can be used to represent arbitrary geometries and has gained its popularity in \rev{biological} fluid mechanics\rev{. For example, interactions between} a very complex body, such as a bluegill sunfish pectoral fin or a false vocal fold, and its surrounding flows have been studied in a two-way coupled manner \rev{via the first method} \citep{zheng2009computational,dong2010computational}.
	This method, with an unstructured surface mesh for the complex boundary, was introduced and implemented in the Pencil Code by \cite{luo2016ghost}. Arbitrary two-dimensional immersed boundaries are represented by many small line-segments. Each line-segment is identified by two vertices as shown in Fig.~\ref{fig:IBM_complex}. The general procedure is still the same as illustrated in Section \ref{subsec:IBM}, other than some special handling of fluid-points and mirror-points around the solid object. 	
	
	The first difference lies in the identification of a given grid point to be a fluid-point or a solid-point. For a circular object, the distance from a given grid point to the center of the circle is calculated, and compared with the radius of the circular object to identify if this grid point is a solid grid point or not, i.e. if it is inside the object or not. This becomes an ineffective method for a complex geometry, where no single radius can be found for the object. In this case, for a given grid point, the closest surface element is detected first. Secondly, the dot product between the closest line-segment's normal vector and the direction vector pointing from the centroid of the closest facet to the given grid point is calculated. The sign of the dot product determines the identification of the grid point. Generally, a negative result indicates a solid-point for the convex boundary shown in Fig.~\ref{subfig:IBM_complx_points}. The treatment of some special cases that may occur during this process is described in \cite{luo2017ghost}. 
	
	After the identification of solid/fluid-points, three layer\rev{s} of ghost points are assigned to construct a six-order central finite-difference stencil as shown in Fig.~\ref{subfig:IBM_complx_mirror}. Following this, a corresponding boundary intercept point is determined for each one of them. The method to detect the boundary intercept points is different from that of the simple circular geometry. First, the vertex closest to a given ghost point is determined. Then, the set of line/surface elements sharing that vertex can be identified and a search is carried out among these elements to find the boundary intercept point (which should lie within the line/surface elements) as shown in Fig.~\ref{subfig:IBM_complx_mirror}. While conceptually simple, the implementation can be very complicated and special attention is needed to find the correct intercept point. Here, we adopt a method based on the robust procedure proposed in \cite{mittal2008versatile}. For details, see \cite{luo2017ghost}. 
	
	Once boundary intercept points are determined for every ghost point, a corresponding mirror-point can be obtained. The mirror-points are set either by symmetry over the solid's line element (corresponding to the way mirror-points are set with a simple geometry, see Fig.~\ref{fig:IBM}), or at a constant distance away from the boundary intercept point. This distance, $ \delta $ in Fig.~\ref{subfig:IBM_complx_mirror}, is typically set to $ \sqrt{2}\Delta x $ for 2D geometries and $ \sqrt{3}\Delta x $ for 3D-geometries, to ensure that every mirror-point is surrounded by fluid-points only. The same interpolation procedure as for the simple circular geometry (i.e., the bi-linear interpolation method) can be adopted for the calculations of the parameters of the mirror-points. An optional way is the inverse distance weight interpolation method \citep{chaudhuri2011use}. \rev{Finally, the flow variables at the ghost-points can be calculated with the aid of the mirror-points and the given boundary conditions at the boundary intercept point by linear interpolation. Three types of boundary conditions, the Dirichlet, Neumann or Robin boundary condition have been implemented, similarly as for the simple circular object discussed in Section 2.2.1. For more details and test of the boundary conditions, the reader is referred to \cite{luo2016ghost}.}
	
	
	This new method can be straightforwardly extended to complex three-dimensional geometries, where triangular surface elements can be adopted to represent the surfaces. Details about the method and its implementation can be found in \cite{luo2017ghost} and will not be repeated here. In \cite{luo2017ghost}, a spatial convergence test indicates that only the bi-linear interpolation procedure can obtain a local second-order accuracy. Systematic validations have also been conducted through calculations of flow past an elliptical cylinder, square cylinder, semi-cylinder, as well as a NACA\rev{0}012 airfoil. 
	\rev{Quantitative comparisons with reported results in literature show that the present method can accurately reproduce the main features of the fluid flow past solid objects with complex geometry, quantified by coefficients such as drag and lift coefficients, Nusselt number and Strouhal number}.

	\begin{figure}
		\centering
		\includegraphics[width=0.5\linewidth]{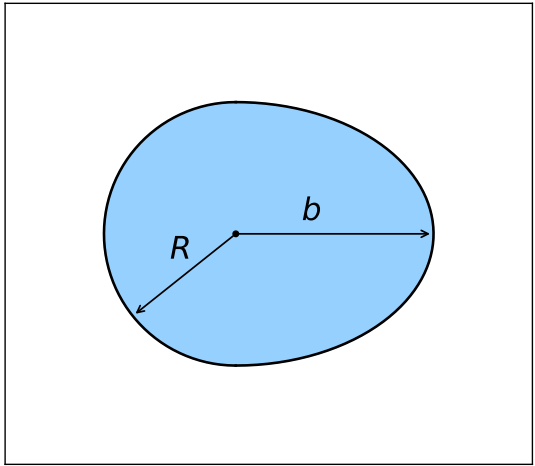}
		\caption{Complex geometry. A combination of a semi-circular cylinder with radius $ R $, and semi-elliptical cylinder with major axis $ b $ and minor axis $ R $. The circular side of the geometry faces the inlet.}
		\label{fig:compx_geometry}
	\end{figure}

	To demonstrate the IBM capabilities for a two-dimensional flow, we have simulated flow past geometries constructed by combining a semi-circle and semi-elliptical cylinder. The geometry is seen in Fig.~\ref{fig:compx_geometry}. The radius of the semi-circle ($ R $) and the major axis of the semi-ellipse ($ b $) can be varied to construct different geometries. Three cases, with $ R/b =$ 2.0, 1.0 (circle), and 0.5, respectively, are considered. For each case, 360 line segments are used to resolve the immersed geometry. Other parameters related to the computational domain are kept consistent with the $Re=100$ case in the grid refinement part of Section 3, except that the solid body is placed at a distance $ 5D $ from the inlet, in the streamwise direction, rather than in the center of the flow domain ($ 10D $ from the inlet).

	\begin{figure}
		\centering
		\subfigure[{$ R/b = 2.0 $}]{%
			\includegraphics[width=0.28\linewidth]{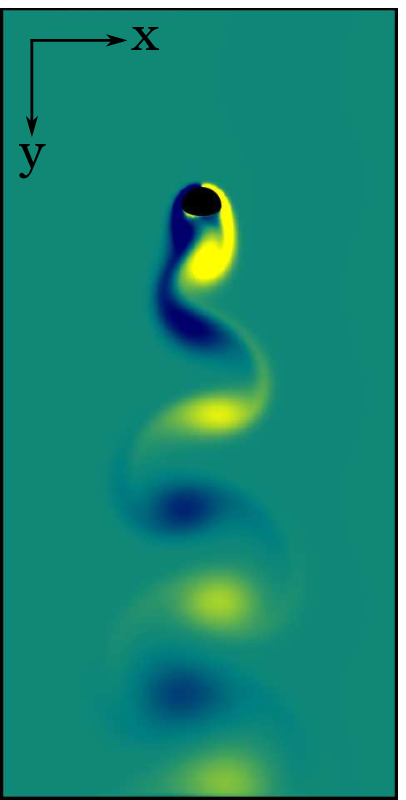}%
			\label{subfig:visuG1}}\qquad
			\centering
		\subfigure[{$ R/b = 1.0 $}]{%
			\includegraphics[width=0.28\linewidth]{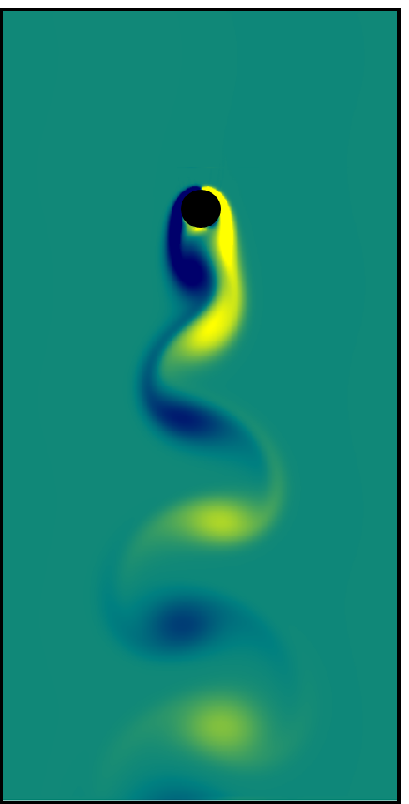}%
			\label{subfig:visuG2}}\qquad
		\subfigure[{$ R/b = 0.5 $}]{%
			\includegraphics[width=0.28\linewidth]{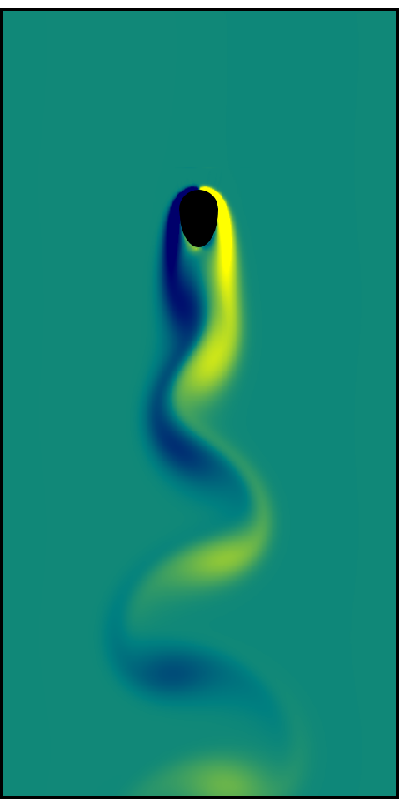}%
			\label{subfig:visuG3}}
		\caption{Flow visualization. Contours of instantaneous vorticity $ \omega_z = \left[\nabla \times \vect{u}\right]_z $ (normal to the view plane) plotted for three different geometries at $ Re=100 $. Inflow at the top of plane.}
		\label{fig:visualization_complex}
	\end{figure}
	
	The vorticity component \rev{$ \omega_z $} normal to the view plane for flow past the three different geometries can be seen in Fig.~\ref{fig:visualization_complex}. It can be seen that as the length $ b $ of the semi-ellipse is increased, the length of the bound vortex increases accordingly. This results in different patterns of von K\'{a}rm\'{a}n vortex streets for each of the three cases. 	
	The corresponding mean drag coefficient, root-mean-square lift coefficient, and Strouhal frequency number are listed in Tab.~\ref{tab:results_complex}. It is obvious that even though the immersed boundary of the third geometry is the longest, the drag force it experiences is the least. 
	This is perhaps not surprising, as the shape of the object is \rev{closer to a streamlined body} for the largest value of $ b $, a shape that is known for low drag.

	\begin{table}
		\centering
		\caption{Comparisons of mean drag coefficient, root-mean-square lift coefficient, and Strouhal number \rev{for} different geometries.}
		\label{tab:results_complex}
		\begin{tabular}{*{4}{c}}
			\hline
			$ R/b $ & $C_D$ & $C_L'$ & $St$\Tstrut\Bstrut \\
			\hline
			2.0 & 1.55 & 0.34	& 0.190\Tstrut   \\
			1.0 & 1.35 & 0.26 & 0.175 \\
			0.5 & 1.17 & 0.11 & 0.168\Bstrut \\
			\hline 
		\end{tabular}
	\end{table}

	\section{Concluding remarks \label{sec:conclusion}}
	In this study we have described and compared the two solid body representations available in the high-order finite-difference code known as the Pencil Code. The two methods, the immersed boundary method and overset grids, are fundamentally different in many aspects. These differences can be summed up as:
	\begin{itemize}
		\item[(a)]
		The ghost-point IBM can be implemented straightforwardly in an existing flow solver, by extending the code without requiring major modifications to the existing solver. Using overset grids requires a more generalized flow solver, able to handle all grids that are overset one another (Cartesian, cylindrical, etc.). This may require a modification of the flow solver itself, when overset grids are first implemented in an existing fluid dynamics code.
		\item[(b)]
		Neither the IBM nor overset grids are mass conserving, as they both rely on interpolation of flow quantities to either mirror-points or non-conforming grid points, respectively. As the interpolation is moved away from the solid surface when overset grids are used, the accuracy loss following from interpolation is expected to have a reduced impact on flow properties directly related to the solid object (boundary layer properties, etc.).
		\item[(c)]
		For a circular cylinder, using the overset grid method in the Pencil Code is far superior to using IBM. Reaching grid independent solutions with IBM required 4.7 and 4 times as many grid points in each direction for $ Re=100 $ and 400, respectively, as compared to the background grid used in the overset grids method. In total, this meant using 18.1 and 14.5 times as many grid points in our two-dimensional simulations with IBM as compared to that with overset grids at these Reynolds numbers. In addition there comes a much stricter limitation on the time step for the fine grid used in the IBM. Such a limitation is only present on the curvilinear grid in the overset grid method, while a 4-5 times as large time step can be used on the coarse background grid.
		\item[(d)]
		The IBM is highly flexible. The implementation in the Pencil Code can handle complex geometries, that is, surfaces where an analytic surface representation is not available. This opens up a large area of research, that cannot be studied with overset grids.
	\end{itemize}
	
	The development of both IBM and overset grids is far from over, and the evolution of these methods in the Pencil Code is destined to continue as long as researchers use the code and implement their own improvements into this open-source software. Perhaps, in time, overset grids can become more flexible, and the high-order accuracy achieved for the simple geometry can be available for more complex shapes (e.g., by using many grids overset one another). Alternatively (or, perhaps, in addition), the accuracy of the IBM implementation may be improved through implementation of stable, high-order interpolation of mirror-points in the vicinity of the solid object. We hope that with this paper, more researchers will be attracted to use the Pencil Code for simulations of flow past solid objects. Only in this way can the advancements of the solid object representations in the software continue, in the spirit of open-source software development. 
	
	\section*{Acknowledgements} 
	We would like to acknowledge that this research is funded by the Research Council of Norway (Norges Forskingsr\aa d) under the FRINATEK Grant [grant number 231444] and by the National Natural Science Foundation of China [No. 51576176]. The research is supported in part by The GrateCFD project [grant 267957/E20], which is funded by: LOGE AB, Statkraft Varme AS, EGE Oslo, Vattenfall AB, Hitachi Zosen Inova AG and Returkraft AS together with the Research Council of Norway through the ENERGIX program. Computational resources were provided by UNINETT Sigma2 AS [project numbers NN9405K, NN2649K].
	
	\appendix
	\section{\rev{Sample cases} \label{app:sample}}
	\rev{To get started with simulations of flow past a cylindrical geometry using the Pencil Code, sample cases that are available with the download of the code \citep{pencilcode} may be useful. From the pencil-code directory, the sample cases can be found in:}
	
	\verb|./samples/2d-tests/cylinder_deposition|
	
	\verb|./samples/2d-tests/cylinder_deposition_ogrid|
	
	\noindent
	\rev{The postfix ogrid denotes the overset grid sample case. To compile and run a sample case, use the commands} \verb|pc_build|, \verb|pc_start| \rev{and} \verb|pc_run|.	
	\rev{Both sample cases are simulations of a particle-laden flow past a cylinder at $ Re=100 $, in which particles may impact and deposit on the cylinder surface. For documentation on the handling of particles and particle deposition with the cylinder represented by IBM and overset grids, the reader is referred to \cite{Haugen2010} and \cite{aarnes2018ogrid}, respectively.}


	\bibliographystyle{tfv}
	\bibliography{./ref}
\end{document}